\documentclass[manuscript,acmsmall,nonacm]{acmart}
\setcopyright{none}
\AtBeginDocument{%
  }

\setcopyright{acmlicensed}
\copyrightyear{2018}
\acmYear{2018}
\acmDOI{XXXXXXX.XXXXXXX}
\acmConference[Conference acronym 'XX]{Make sure to enter the correct
  conference title from your rights confirmation email}{June 03--05,
  2018}{Woodstock, NY}
\acmISBN{978-1-4503-XXXX-X/2018/06}

\usepackage{booktabs}
\usepackage{tabularx}
\usepackage{array}

\begin{document}

\title{Making Sense of Audience Change: Causal Attribution and Boundary Management on Xiaohongshu}


\author{Erin Zhou}
\email{zzrin@uw.edu}
\affiliation{%
  \institution{University of Washington}
  \city{Seattle}
  \state{Washington}
  \country{USA}
}

\author{Kyrie Zhou}
\email{kyrie.zhou@utsa.edu}
\affiliation{%
  \institution{University of Texas at San Antonio}
  \city{San Antonio}
  \state{Texas}
  \country{USA}
}

\renewcommand{\shortauthors}{Zhou and Zhou}

\begin{abstract}
Recommendation-driven platforms can reshape not only what users see, but also who becomes visible and how community boundaries are negotiated. We examine Xiaohongshu discussions surrounding a perceived World Cup-related audience influx to understand how users attributed community change and how those explanations related to boundary-management strategies. We analyzed 262 comments from eight high-engagement threads using researcher-led thematic analysis, structured coding, and permutation-based association analysis. Users described personalization as both social separation and a boundary that could fail, interpreted platform growth through audience acquisition and commercialization, and contested whether gender should define community membership. Among 70 comments with explicit attributions and strategies, attribution and strategy were strongly associated (Cram\'er's $V=.583$): algorithmic attributions concentrated in technical boundary management, while social-actor attributions more often aligned with social enforcement or normative socialization. These findings connect causal sensemaking with boundary management and show how audience restructuring can create governance consequences on recommendation-driven platforms.
\end{abstract}




\maketitle

\section{Introduction}

Recommendation-driven platforms can change not only what users see but also who becomes visible and which participation norms come into contact. In June 2026, Xiaohongshu launched a World Cup channel. In the public discussions we examine, users often interpreted the event as an effort to attract or retain new audiences, especially male users. Because Xiaohongshu is organized heavily through personalized recommendation rather than a single platform-wide membership boundary, perceived audience change appeared through feed exposure, encounters among users, renewed norm debate, and competing claims about belonging.

Prior HCI research provides several pieces for understanding this problem. Studies of rapid newcomer influx and group mergers show that participation changes can make existing norms newly consequential and create governance work in bounded communities \cite{kiene2016eternal,kiene2018mergers}. Research on algorithmic folk theories shows that users develop explanations for opaque platform systems and that those explanations can relate to responses to perceived technical change \cite{devito2017algorithms}. Recommendation can also organize social relationships: Wang et al. describe an ``algorithmically woven community'' on Xiaohongshu in which recommendation connected users with similar experiences \cite{wang2023woven}. On Xiaohongshu, Wan et al. show women re-appropriating hashtags to reduce unwanted male audience exposure amid concerns about a shifting user base, harassment, and inadequate safety tools \cite{wan2025hashtag}. Together, this work links audience change, recommendation, sensemaking, and boundary management.

A remaining question is how these processes interact when the object of sensemaking is perceived change in a broader platform community. Public explanations of sociotechnical outcomes can distribute responsibility across algorithms, platforms, groups, and individuals \cite{zhang2025beyond}, while studies of ``TikTok refugees'' on Xiaohongshu document negotiation over newcomer participation, visibility, cultural boundaries, and platform norms \cite{yuan2025internet,xiao2025nolongero}. We know less about how users publicly explain a perceived platform-associated audience shift when recommendation mediates social encounters, or whether different causal locations are systematically associated with different ways of negotiating community boundaries. This is especially important when the boundary itself is informal and contested, because users may disagree over whether gender, conduct, interests, platform history, or everyday use should define membership.

We examine public Xiaohongshu discussions surrounding the perceived World Cup related audience influx. We purposively selected eight high-engagement threads concerning the World Cup, male-user growth, and perceived community change. From 808 deduplicated public comments and replies, a conservative relevance audit produced a final analytic corpus of 262 comments. We combined researcher-led thematic analysis with structured coding of stance, causal attribution, and boundary-management strategy, then used permutation-based association analysis to examine whether attributions and strategies co-occurred systematically. We ask:

\begin{enumerate}
\item[\textbf{RQ1.}] How did users participating in public discussions around the World Cup related audience influx make sense of and attribute perceived community change?

\item[\textbf{RQ2.}] What strategies did users articulate or enact to negotiate, defend, or adapt perceived community boundaries in response to the influx?
\end{enumerate}

Users described personalization as both separation between social and interest circles and a boundary that could fail when unwanted content remained visible. They interpreted the World Cup investment through audience acquisition, retention, and commercialization, debated norms of participation, and disagreed over whether gender should define membership. Among the 70 comments containing both an explicit attribution and strategy, attribution and strategy were strongly associated ($p < .001$, Cramer's $V = .583$): algorithmic attributions concentrated in technical boundary management, whereas social-actor attributions more often co-occurred with social enforcement or normative socialization. Attribution was not significantly associated with evaluative stance ($p = .182$, Cramer's $V = .200$). The clearest pattern therefore concerned how perceived problem location aligned with articulated intervention site.

This paper makes three contributions to HCI:
\begin{itemize}
\item We provide an event-focused account of how users interpreted and negotiated perceived community change on a recommendation-driven platform without a formal platform-wide membership boundary.
\item We connect causal sensemaking with boundary management through a systematic, observational association between perceived problem location and articulated intervention site.
\item We refine accounts of community and governance by showing how audience acquisition can create governance consequences, personalization can be experienced as social boundary infrastructure, and gendered membership can remain meaningful while its legitimacy as a membership criterion is contested.
\end{itemize}

\section{Related Work}

\subsection{Community Norms and Gendered Boundary Negotiation}
\label{sec:rw-community-boundaries}

Online communities are sustained partly through norms that define appropriate participation and can function as governance even without formal platform rules. Dym and Fiesler show that norms supporting privacy and safety in transformative fandom become vulnerable when subcommunities hold conflicting expectations, outsiders misunderstand local practices, or platform design fails to support community needs \cite{dym2020socialnorm}. Beadle et al. similarly show that implicit norms in non-binary Reddit communities are created, contested, and reinforced through everyday interaction \cite{beadle2025implicit}, while large-scale Reddit moderation research finds that norms vary across macro, meso, and micro levels \cite{chandrasekharan2018hiddenrules}. Boundary maintenance can also require explicit sociotechnical work. Wu et al.'s study of racist attacks in city-based Reddit communities shows how moderation practices established and defended institutional, cultural, and geographical boundaries when community norms were challenged \cite{wu2024boundaries}. Together, these studies treat community boundaries as ongoing social and technical accomplishments rather than fixed properties.

Governance also shapes which identities and practices become associated with a community. Das et al.'s trace ethnography of Bengali Quora shows how sociotechnical governance can privilege some linguistic, national, and religious identities while marginalizing others, contributing to platform identity \cite{das2021jol}. Identity-based boundaries may also be explicit: Ammari et al.'s interviews with moderators of women-only Facebook groups show how membership is actively maintained through moderation and platform affordances \cite{ammari2022moderation}. Yet which identities should organize participation can itself be disputed. Qin et al. identify competing gendered and gender-blind framings in Chinese social-media discourse, while Deng et al. examine discursive strategies in public gender debate \cite{qin2024genderblindness,deng2024persuasion}. Haapoja et al. further show that appropriate participation can be evaluated through roles and conduct \cite{haapoja2025ideal}. Community identity is therefore produced through governance, discourse, and competing claims about which distinctions and practices should matter.

Research on Xiaohongshu provides a more specific context for gendered boundary work. Gu et al. show how \#StopMenstrualShaming supported feminist critique, solidarity, and community formation while commercial dynamics shaped participation \cite{gu2024commerce}. Chang similarly shows feminist creators adjusting language, content, and platform practices to maintain fluid semi-public spaces under censorship and misogynistic hostility \cite{chang2025safespaces}. The closest precedent is Wan et al.'s mixed-methods study of hashtag re-appropriation \cite{wan2025hashtag}. Analyzing 5,800 posts and interviewing 24 active users, they show women repurposing hashtags such as \#Baby~Supplemental~Food to influence recommendation and reduce unwanted male audience exposure. Participants connected this practice to a shifting user base, gender-based harassment, dissatisfaction with blocking and reporting, and perceived changes in the platform's orientation toward male users.

Existing work thus establishes that identity-based communities actively maintain norms and boundaries and that Xiaohongshu users have developed recommendation-mediated audience-control practices. A different problem arises when the relevant boundary is informal and its legitimacy is itself disputed. Research has said less about how users publicly negotiate membership in a broader platform community during perceived audience change when gender, conduct, interests, and historical platform culture offer competing bases for belonging. We examine boundary management in this setting as negotiation over both participation practices and the criteria through which membership is understood.

\subsection{Audience Management and Algorithmic Sensemaking on Recommendation-Driven Platforms}
\label{sec:rw-algorithmic-audiences}

Managing uncertain audiences has long been a central problem of social media participation. Marwick and boyd describe Twitter users constructing imagined audiences and adapting self-presentation when normally separate social contexts become difficult to distinguish \cite{marwick2011tweet}. Davis and Jurgenson theorize context collapse as involving both architectures and user practices that bring contexts together or help keep them apart \cite{davis2014context}. Empirical HCI work likewise shows substantial uncertainty about who actually encounters content: audience size and composition can be difficult to estimate, and platform affordances may expand audiences while revealing little about who is present \cite{bernstein2013invisible,devito2017platforms}. Audience management therefore depends on partial and platform-mediated understandings of visibility.

Algorithmic curation adds uncertainty about how platforms select content and audiences. Rader and Gray found varied causal beliefs about Facebook News Feed curation \cite{rader2015beliefs}; Eslami et al. identified multiple folk theories and showed that users revised explanations when aspects of filtering became visible \cite{eslami2016folk}. DeVito et al. later showed that folk theories draw on multiple sources and remain complex and malleable \cite{devito2018folk}. Adaptive folk theorization further demonstrates how users revise theories as platforms change and incorporate broader perceptions of platform ``spirit'' \cite{devito2021adaptive}. Algorithmic sensemaking is thus an ongoing interpretation of both technical mechanisms and users' relationship with the platform.

A central question is how explanations relate to action. During \#RIPTwitter, more detailed folk theories tended to accompany more specific reactions to anticipated algorithmic change, without establishing causality \cite{devito2017algorithms}. Mayworm et al. similarly show moderation folk theories informing changes in vocabulary, reduced participation, and platform exit \cite{mayworm2024moderation}. Responses can also be collective: users develop strategies for controlling algorithms \cite{burrell2019control}, circulate ``algorithmic gossip'' about visibility \cite{bishop2019gossip}, and coordinate large-scale attempts to influence recommendation \cite{xiao2025influence}. These studies connect algorithmic sensemaking with perceived sites of control and individual or collective adaptation.

Personalization also shapes social relationships. Karizat et al. show TikTok users theorizing how algorithms treat personal and marginalized identities and engaging in algorithmic resistance \cite{karizat2021algorithmic}; Lee et al. show personalization shaping self-understanding and orientation toward others \cite{lee2022crystal}. On Xiaohongshu, Wang et al. describe an ``algorithmically woven community'' in which recommendation connected users with similar mental-health experiences across a loosely knit network \cite{wang2023woven}. On Zhihu, some long-term gay male users understood filtering as protective because it reduced hostile encounters and supported relatively exclusive networks \cite{zhao2023filter}. Recommendation can therefore participate in producing both social proximity and separation.

Wan et al. provide the closest Xiaohongshu-specific account of users intervening in this process. Women adopted hashtags they believed would be less appealing to male audiences to influence recommendation and reduce unwanted exposure; the tactic changed as those hashtags gained broader visibility \cite{wan2025hashtag}. Research also moves beyond explanations centered on algorithms alone. Zhang and Yang's comparative study of information-homogeneity discourse identifies responsibility distributed across groups, individuals, political actors, platforms, and algorithms, with differences between Chinese and U.S. contexts \cite{zhang2025beyond}. Multi-actor attribution is therefore not unique to our case. We extend this perspective to perceived change in a broader platform community and ask whether different causal framings are systematically associated with different forms of articulated boundary management.

\subsection{Audience Change, Newcomer Influx, and Platform Governance}
\label{sec:rw-platform-growth}

Changes in participation can reshape communities even when formal rules remain unchanged. Kiene et al.'s study of NoSleep examined sudden membership growth after Reddit automatically subscribed new users; the community maintained norms through coordinated administration, member enforcement, and technical mechanisms \cite{kiene2016eternal}. Their later study of World of Warcraft guild mergers shows that outcomes also depend on newcomers' established relationships and practices, cultural compatibility, socialization, and solidarity \cite{kiene2018mergers}. These cases demonstrate that membership change alters the conditions under which norms are reproduced.

Community change also occurs through migration and platform architecture. Fiesler and Dym show transformative fandom migration being shaped by technical design, policies, social relationships, and community practices, sometimes fragmenting ties or content \cite{fiesler2020moving}. Zhang et al.'s Form-From design space distinguishes systems organized through spaces, networks, or broader commons, helping compare formally bounded groups with recommendation-driven settings where community-like interaction can emerge without explicit membership \cite{zhang2024formfrom}. Research on moderator engagement and self-moderation further shows community-level actors developing norms and governance \cite{seering2019moderator,seering2020selfmoderation}, while Jhaver et al. locate authority across platform operators, intermediate governance structures, and end users \cite{jhaver2023decentralizing}.

Recent Xiaohongshu research makes audience change especially salient. Wan et al. link audience-control practices to perceptions of a shifting user base and greater platform orientation toward male users \cite{wan2025hashtag}. Studies of the 2025 ``TikTok refugee'' migration identify newcomer communication strategies and negotiations over visibility, cultural and linguistic boundaries, platform norms, and algorithmic attention \cite{yuan2025internet,xiao2025nolongero}. This literature explains newcomer influx, migration, architecture, and governance, but less often examines public sensemaking of a perceived audience shift associated by users with a platform-side content and growth intervention. Our case focuses on how existing users distributed responsibility across platform, commercial, algorithmic, and social actors while negotiating an informal community boundary whose meaning was contested.

\section{Methods}
\label{sec:methods}

Our analysis combined researcher-led thematic analysis with structured operational coding and quantitative association analysis. We first constructed an event focused corpus of public Xiaohongshu discussions, screened comments for relevance to the research questions, and developed qualitative themes through iterative analysis. We then applied a separate operational codebook to examine stance, attribution, and articulated boundary management strategies.

\begin{figure}[t]
    \centering
    \includegraphics[width=\linewidth]{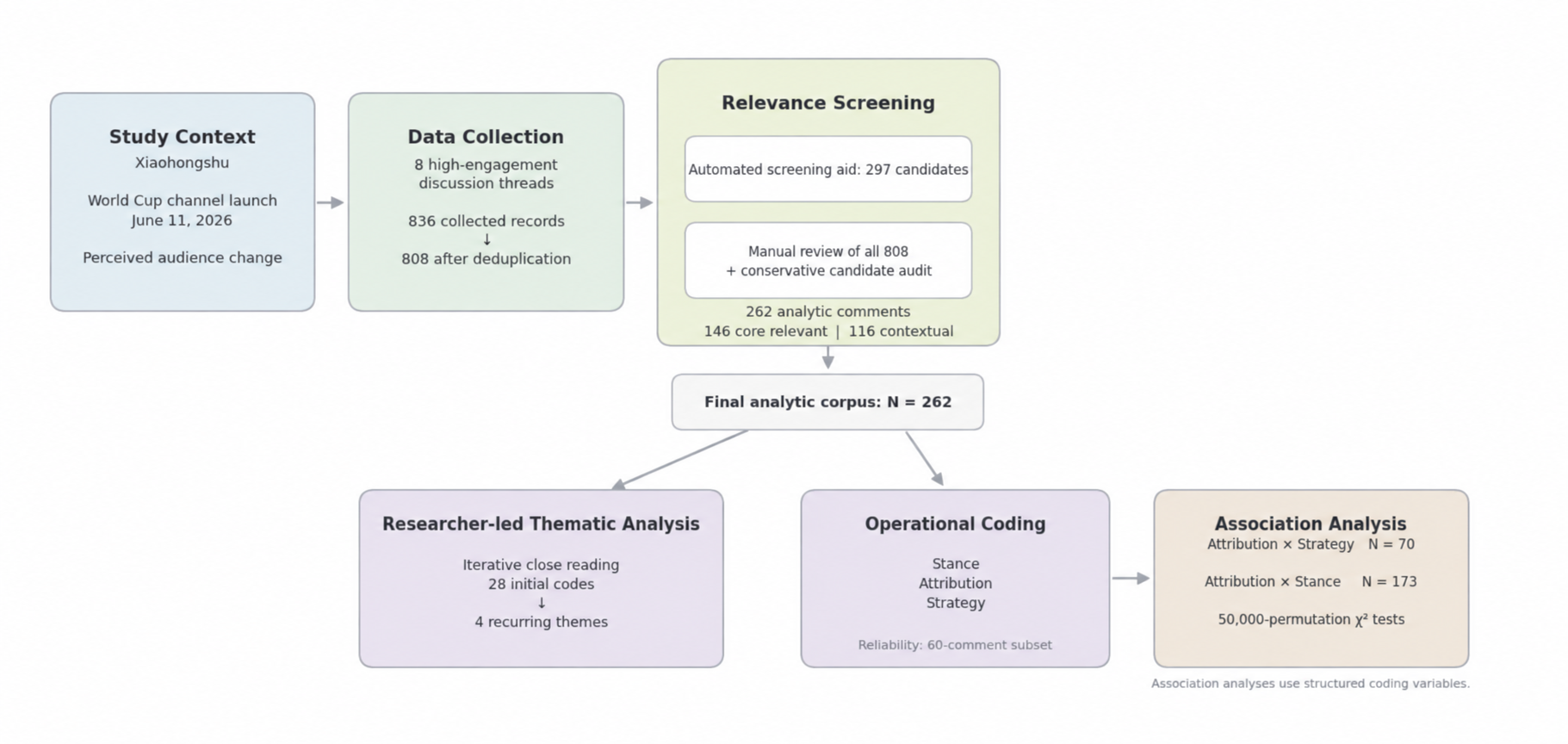}
    \caption{
    Overview of corpus construction and analysis. The study began with eight high-engagement Xiaohongshu discussion threads surrounding the perceived World Cup-related audience influx. After collecting 836 comment records and deduplicating them to 808 comments, relevance screening produced a final analytic corpus of 262 comments. The final corpus was analyzed through researcher-led thematic analysis and separate structured operational coding. The structured coding variables were then used in the quantitative association analyses.
    }
    \Description{
    A study overview diagram. Public Xiaohongshu discussions surrounding the World Cup-related audience change were collected from eight high-engagement threads, producing 836 comment records and 808 comments after deduplication. Screening identified 297 candidate comments, while the first author manually reviewed all 808 comments. The final analytic corpus contained 262 comments, including 146 core relevant and 116 contextual comments. The final corpus was analyzed through researcher-led thematic analysis and operational coding of stance, attribution, and strategy. Operational coding supported reliability assessment and association analyses of attribution by strategy and attribution by stance.
    }
    \label{fig:study-overview}
\end{figure}

\subsection{Study Context and Data Collection}
\label{sec:data-collection}

We examined public Xiaohongshu discussions surrounding perceived changes in the platform's audience following its investment in World Cup content. Data collection began on June 11, 2026, when Xiaohongshu launched its World Cup channel. We used this event as a focal case for examining how users interpreted and responded to a visible period of audience change on a recommendation-driven platform.

We identified source posts through searches on Xiaohongshu using combinations of Chinese search terms corresponding to ``male users,'' ``World Cup,'' ``increase in male users,'' and ``Xiaohongshu.'' We first screened search results for substantive relevance. Posts were eligible when their main discussion concerned the World Cup-related audience influx, increased male participation, or perceived changes in the Xiaohongshu community. Among these eligible results, we purposively selected eight posts with comparatively high visible engagement, assessed from their comment and like activity. The source set was designed to capture active public discourse around the focal event. We do not treat it as representative of Xiaohongshu activity as a whole.

The first author wrote a custom script to collect publicly visible comments and replies available from the eight source posts at the time of collection. The collection produced 836 comment records. We removed exact duplicate comment texts occurring within the same source thread, resulting in a deduplicated corpus of 808 comments. Each comment retained a source identifier so that source specific patterns could be examined during analysis.

\begin{table}[htbp]
    \centering
    \caption{Overview of the eight source discussions and their contributions to the analytic corpus.}
    \label{tab:source-overview}
    \small
    \setlength{\tabcolsep}{5pt}
    \renewcommand{\arraystretch}{1.05}

    \begin{tabularx}{\linewidth}{
        @{}
        l
        >{\raggedright\arraybackslash}X
        r
        r
        @{}
    }
        \toprule
        Source & Focal discussion & Collected & Analytic \\
        \midrule
        D01 & Increase in male users & 90 & 31 \\
        D02 & World Cup investment & 191 & 76 \\
        D03 & Attracting male users & 107 & 35 \\
        D04 & World Cup and community change & 30 & 14 \\
        D05 & Arrival of male users & 17 & 5 \\
        D06 & World Cup audience expansion & 42 & 18 \\
        D07 & Male users and community change & 65 & 27 \\
        D08 & Male user growth and migration & 266 & 56 \\
        \midrule
        \textbf{Total} & & \textbf{808} & \textbf{262} \\
        \bottomrule
    \end{tabularx}

    \vspace{3pt}

    \parbox{\linewidth}{\footnotesize
    Collected counts refer to deduplicated publicly visible comments and replies. Analytic counts reflect the relevance screening described in Section~\ref{sec:corpus-screening}.}
\end{table}

\subsection{Research Ethics and Data Protection}

The university's Institutional Review Board (IRB) oversees research involving human participants. This study analyzed publicly available social media posts without interaction or intervention with users. Under applicable institutional requirements, formal IRB review was not required for this research.
We collected only information necessary for the analysis and did not retain usernames, profile images, account identifiers, or other profile-level information. We did not infer demographic characteristics from usernames, avatars, writing style, or profile content.

To reduce the risk that individual users could be identified through text searches, quotations were translated from Chinese into English and lightly paraphrased while preserving their substantive meaning. Source identifiers used in the paper refer only to anonymized discussion threads and comments and cannot be linked to public account identifiers through the reported materials. The dataset containing verbatim comments will not be publicly released because public availability does not eliminate the potential privacy risks associated with aggregating and redistributing user-generated content.

\subsection{Corpus Construction and Relevance Screening}
\label{sec:corpus-screening}

Comments appearing in a relevant source thread did not automatically qualify for analysis. We therefore determined relevance at the comment level. Core relevant comments directly addressed the World Cup related audience shift, perceived changes in male or newcomer participation, community change, explanations for these changes, or responses to them. Contextual comments contributed substantive background about platform strategy, recommendation, community norms, gendered audience composition, or related mechanisms. Comments that did not materially contribute to either research question were classified as off topic.

We initially used a codebook-based few-shot classification procedure to organize the 808 deduplicated comments. This pass identified 297 comments as potentially core relevant or contextual. The automated classifications served only as a screening aid. The first author subsequently reviewed all 808 comments using the same relevance criteria, including comments initially classified as off-topic. Thread membership alone was not treated as evidence of relevance.

A second audit applied the inclusion criteria conservatively to the 297 candidate comments. Thirty five comments were removed because their content did not materially contribute to either research question. The manual review of the initially excluded comments identified no additional comments for inclusion. The final analytic corpus contained 262 comments, including 146 core relevant and 116 contextual comments. All subsequent analyses reported in the paper used this final corpus.

\subsection{Researcher Led Thematic Analysis}
\label{sec:thematic-analysis}

We conducted a researcher led thematic analysis to examine recurring ways users understood and responded to perceived community change. The first author conducted the qualitative analysis and is a native Chinese speaker familiar with Xiaohongshu and its interactional conventions. This familiarity supported interpretation of colloquial language, platform specific expressions, and context dependent references. Iterative review and explicit attention to contradictory cases were used to challenge initial interpretations.

The unit of analysis was the individual comment. Comments were also read within their source thread when reply context or earlier discussion was needed for interpretation. The thematic analysis was analytically distinct from the operational classification described in Section~\ref{sec:operational-coding}. Automated relevance, stance, attribution, and strategy labels were not treated as themes.

Theme development began with close reading of the 297 candidate comments and analytic notes concerning platform growth, recommendation, community norms, responses to unwanted encounters, and contested understandings of membership. Comments could receive multiple initial codes. Repeated comparison produced 28 initial codes, which were then clustered around four broader areas concerning algorithmic boundaries, platform growth, community norms and boundary management, and contested gendered membership.

The first author iteratively reviewed candidate themes for internal coherence and distinction from one another. Contradictory evidence remained part of this process. Comments describing successful personalization constrained an account focused only on algorithmic failure. Positive evaluations of audience expansion constrained a uniformly critical interpretation of platform growth. Conduct-based accommodation complicated an exclusion centered account of boundary management. Comments rejecting gender as a criterion for belonging also constrained a simple gender conflict interpretation. We additionally rejected a proposed escalation narrative between different boundary management strategies because the comment data did not establish temporal ordering.

After the final relevance audit reduced the corpus from 297 to 262 comments, the first author rechecked the thematic coding against the final corpus. All four themes remained represented across multiple source threads, and their coverage and negative cases were recalculated using the 262 comment corpus. The removal of off topic comments did not require adding, removing, or redefining the four themes. Quotations reported in the paper were translated from Chinese into English by the first author.

\subsection{Operational Coding}
\label{sec:operational-coding}

Alongside the thematic analysis, the first author applied a structured operational codebook to the final 262 comments. This coding served a separate analytic purpose. Themes captured broader patterns across the discourse, while the operational labels provided mutually exclusive variables for descriptive summaries, reliability assessment, and the quantitative association analyses.

Stance captured evaluation of the focal audience or community change and was coded as \textit{negative}, \textit{neutral}, \textit{positive}, or \textit{unclear}. Attribution captured the primary causal explanation explicitly expressed in a comment and included \textit{platform}, \textit{male users}, \textit{algorithm}, \textit{commercialization}, \textit{community}, \textit{other}, and \textit{none}. A topical reference alone was insufficient for assigning an attribution.

Strategy captured the primary boundary management or coping action articulated or enacted in the comment. The final categories were \textit{algorithmic separation}, \textit{content filtering}, \textit{normative accommodation}, \textit{norm education}, \textit{confrontation}, \textit{block and report}, \textit{gatekeeping}, \textit{disengagement}, and \textit{none}. When more than one attribution or strategy was present, we coded the mechanism foregrounded in the comment. The \textit{none} category indicates that no explicit attribution or strategy was articulated in that comment. It does not imply that the commenter held no explanation or used no strategy outside the observed text.

Coding decisions were conservative when textual evidence was insufficient. We did not infer commenter gender from usernames, avatars, writing style, interests, or other profile cues. Gender specific interpretation was limited to explicit self identification or explicit collective positioning in the comment.

\subsection{Reliability and Robustness Checks}
\label{sec:reliability}

We assessed the reproducibility of the operational codebook through independent coding by a second author. Sixty comments were purposively selected from the final corpus, with less frequent strategy categories oversampled to ensure their representation. The subset was designed for reliability assessment and does not represent category prevalence. The second author independently applied the frozen operational codebook without using the first author's labels or model predictions as coding input.

We calculated raw agreement and Cohen's $\kappa$ separately for relevance, stance, attribution, fine grained strategy, and the broader strategy families used in the quantitative analysis. Raw agreement and Cohen's $\kappa$ were .983 and .963 for relevance, .983 and .971 for stance, .950 and .937 for attribution, .967 and .962 for fine grained strategy, and .967 and .957 for broad strategy. Because the reliability subset was drawn from the final analytic corpus, the relevance comparison assessed the distinction between core relevant and contextual comments. These statistics evaluate the reproducibility of the operational categories and were not used to evaluate the interpretive themes.

We also examined sensitivity to the eight source threads. For the primary attribution and strategy analysis, we repeated the analysis eight times while omitting one source thread in each iteration. This check assessed whether the observed association was disproportionately dependent on a single discussion source.

We additionally evaluated the frozen few-shot classifier on a held-out set of 190 comments. Macro F1 was .606 for attribution, .660 for fine-grained strategy, and .719 for the broader strategy families. Performance varied substantially across individual labels, particularly for infrequent norm-oriented strategies. The frozen relevance and stance evaluation preceded the final revisions to those two constructs, so those scores do not validate the final relevance and stance definitions. Model predictions were therefore used as supplementary validation and workflow support and were not the sole basis for the final analyses.

\subsection{Quantitative Association Analysis}
\label{sec:quant-association}

We examined whether coded explanations of perceived community change were associated with articulated boundary management strategies. Several fine grained categories contained few observations, so we combined conceptually related labels into broader families based on their coding definitions to reduce sparse cells and support an interpretable contingency analysis.

For attribution, \textit{algorithm} remained a separate family. \textit{Male users} and \textit{community} were grouped as social actor attributions because both located the perceived source of change in users or social interaction. \textit{Platform} and \textit{commercialization} were grouped as structural platform attributions because both located the explanation in platform decisions, growth, or commercial incentives. Comments coded as \textit{none} or \textit{other} were excluded from this analysis.

For strategy, \textit{algorithmic separation} and \textit{content filtering} were grouped as technical boundary management. \textit{Normative accommodation} and \textit{norm education} formed normative socialization. \textit{Confrontation}, \textit{block and report}, and \textit{gatekeeping} formed social enforcement. \textit{Disengagement} formed withdrawal. Comments coded as \textit{none} were excluded from the association analysis.

The primary analysis included 70 comments containing both an explicit attribution covered by the three attribution families and an explicit strategy covered by the four strategy families. Because the resulting contingency table contained sparse cells, we evaluated the Pearson $\chi^{2}$ statistic using 50,000 permutations and report Cram\'er's $V$ as the measure of association strength. This analysis tests co occurrence between coded attributions and strategies within comments and does not establish causal relationships.

As a secondary analysis, we examined attribution and evaluative stance among the 173 comments containing an explicit attribution. We again used a permutation-based $\chi^{2}$ test with 50,000 permutations and report Cram\'er's $V$. This analysis examined whether attribution categories were also associated with variation in how users evaluated the perceived audience change.


\section{Findings}
\label{sec:findings}

Our analysis identified four recurring patterns in how users made sense of and responded to the perceived World Cup related audience influx on Xiaohongshu. Users described recommendation as shaping whether different audiences remained within relatively separate content environments or became unexpectedly visible to one another. They also interpreted the World Cup investment as part of a broader effort to expand Xiaohongshu's audience, often connecting this growth to male users, commercialization, and platform development. At the interactional level, perceived changes in participation brought existing expectations about appropriate conduct to the foreground and prompted users to articulate different ways of managing community boundaries. At the same time, users disagreed over whether Xiaohongshu should be understood as a gendered community and whether gender itself should determine belonging.

These themes overlapped within individual comments. A user could, for example, describe audience growth as a platform strategy while also discussing recommendation, gendered participation, or a preferred response. We therefore use the themes to describe recurring patterns across the discourse instead of mutually exclusive groups of comments. We conclude the Findings with a quantitative analysis of the relationship between users' coded causal attributions and the boundary management strategies they articulated.

\subsection{Personalization as Separation and Its Breakdown}
\label{sec:findings-personalization}

Recommendation was central to how users explained their everyday encounters with different content and audiences. This theme appeared in 68 comments across five of the eight source threads. Users described personalization in two contrasting ways. Some treated Xiaohongshu's highly personalized feeds as a mechanism that allowed different interests and social circles to coexist with limited contact. Others described moments when this separation broke down, especially when World Cup content or contentious discussions continued to appear after users attempted to remove them.

\subsubsection{Personalized feeds kept different circles apart}

Several users described Xiaohongshu as having particularly strong personalized content environments. For these users, a platform wide change in audience composition did not automatically mean that everyone would encounter the new audience in their own feed. One commenter explained,

\begin{quote}
\small
``It actually does not affect me much. Xiaohongshu has a very strong information bubble. After you open one or two posts, your homepage changes completely, so different circles do not affect one another that much.''

\hfill\textit{[D03, C1]}
\end{quote}

The comment presents the personalized homepage as a form of separation between interest based circles. The user recognized that different populations might exist on Xiaohongshu while expecting limited interaction between them. Audience growth therefore appeared less disruptive when the feed continued to organize users into distinct content environments.

Other comments described this separation through differences between people's homepages. One user compared their feed with a friend's and wrote,

\begin{quote}
\small
``Some things get recommended repeatedly after you click them once, while things you never open will basically never be recommended. My friend's homepage and mine are completely different. Hers is makeup, outfits, and travel. Mine is cats, dogs, anime, merchandise, and other things. I have never seen the suggestive content people talk about.''

\hfill\textit{[D08, C31]}
\end{quote}

Here, personalization was visible through comparison. The commenter did not describe Xiaohongshu as having one shared content environment. The same platform appeared differently to two users because their recommendation histories produced distinct everyday experiences. Several comments used the term ``information bubble'' positively in this context. One simply wrote, ``It is fine. I have my own information bubble'' [D01, C81]. Another argued that people with very different interests could remain on the same platform because ``men and women often do not even see each other'' [D01, C85].

These accounts also help explain why some users reported seeing little gender conflict despite participating on a platform where other commenters described it as pervasive. One explicitly male commenter described using Xiaohongshu extensively without frequently encountering antagonistic posts.

\begin{quote}
\small
``I am a man, and I actually rarely see confrontational posts. Xiaohongshu's algorithm is very powerful. Once I click `not interested,' I basically do not see that kind of content again. Xiaohongshu is also like an encyclopedia for my daily life.''

\hfill\textit{[D03, C34]}
\end{quote}

The same platform could therefore be experienced as highly contentious by one user and relatively calm by another. Some commenters explained this difference through recommendation itself.

\subsubsection{Users tried to repair unwanted exposure when separation failed}

The perceived effectiveness of personalization was far from consistent. A substantial set of comments described active attempts to remove World Cup content or other unwanted material. Users mentioned clicking ``not interested,'' changing content preferences, adding blocked terms, blocking accounts, and deliberately searching for preferred topics in an effort to retrain their feeds.

One commenter described repeated failure despite extensive feedback.

\begin{quote}
\small
``You can really see Xiaohongshu's anxiety. I have clicked `not interested' at least fifty times, and it still keeps recommending World Cup content. It is so annoying.''

\hfill\textit{[D02, C185]}
\end{quote}

Other comments described the same pattern in less extreme terms. One user said they clicked ``not interested'' every day and still received World Cup recommendations every day [D02, C66]. Another wrote that they had no interest in football, had marked the content as unwanted many times, and eventually began blocking accounts because the recommendations continued [D02, C118].

These experiences changed what filtering controls meant to users. Clicking ``not interested'' was not simply a preference signal. It was an attempt to restore a feed in which unwanted topics remained outside the user's regular experience. Frustration emerged when the system appeared unable to maintain that separation.

Some users also perceived different topics as being learned at different speeds. One commenter contrasted ordinary interests with contentious gender related content.

\begin{quote}
\small
``Xiaohongshu's recommendation logic feels a little aggressive. If I actively search for anime content, I need to browse several posts before the platform remembers that I like anime. But once I search for gender conflict related content, the platform remembers immediately.''

\hfill\textit{[D08, C51]}
\end{quote}

This comment does not establish how Xiaohongshu's recommendation system technically prioritizes content. It shows that the user interpreted repeated exposure as evidence that controversial topics were especially easy for the system to learn. Similar comments connected conflict, argument, or provocative material with high visibility.

The disagreement between these accounts is an important part of the theme. Some users found a single ``not interested'' action sufficient, while others described repeated feedback as ineffective. One commenter summarized the latter frustration by saying that new accounts quickly encountered conflict content and that requests for controls over tags or user groups had gone unanswered [D03, C71]. Users therefore did not describe a uniformly successful or unsuccessful recommender. They described a system on which they often relied for separation, while disagreeing sharply about how dependable that separation was.

\subsection{World Cup Investment and Audience Growth}
\label{sec:findings-growth}

A second pattern concerned how users explained the audience change itself. This theme appeared in 82 comments across seven source threads. Users frequently interpreted the World Cup investment as a mechanism for acquiring or retaining users. Male users were especially prominent in these explanations because commenters viewed Xiaohongshu's existing audience as female skewed and the World Cup as an opportunity to reach a population that the platform had previously attracted less successfully.

\subsubsection{The World Cup was understood as an audience acquisition event}

For some commenters, the link between the World Cup and audience acquisition appeared straightforward. One user wrote,

\begin{quote}
\small
``A substantial share of men still represents a growth market for the platform. It is that simple.''

\hfill\textit{[D02, C187]}
\end{quote}

Other comments pointed to concrete joining behavior. One commenter said that the platform had ``attracted many male customers'' and added that they personally would not have downloaded Xiaohongshu without the World Cup [D02, C72]. Another simply observed, ``My younger brother downloaded it just for the World Cup'' [D01, C65].

These comments treated media content as an entry point into the platform. The event could bring users who had little previous interest in Xiaohongshu into the app for a specific purpose. Some commenters distinguished these event driven users from dedicated football fans. One argued that ``most are not really football fans, they just like watching the World Cup,'' and suggested that this broader audience could be larger than the population of regular football followers [D02, C131].

Users also questioned whether these new audiences would remain after the event. One commenter predicted,

\begin{quote}
\small
``Xiaohongshu looks popular because people can watch the World Cup there. Once the World Cup is over, at least half of them will probably leave.''

\hfill\textit{[D02, C132]}
\end{quote}

The possibility of temporary participation shaped some explanations of the platform's strategy. Audience growth was discussed in terms of both acquisition and retention. This was especially visible in comments connecting the World Cup to earlier periods of user migration from other platforms.

\subsubsection{Users connected audience growth with commercialization}

Many commenters placed the World Cup within a wider commercial narrative. References to traffic, advertising, user growth, valuation, ecommerce, and a possible public offering appeared across multiple source threads. A concise version of this explanation was,

\begin{quote}
\small
``They are preparing to go public and need more growth, so they bought the World Cup rights.''

\hfill\textit{[D01, C26]}
\end{quote}

Other comments gave a more elaborate account. One user argued that Xiaohongshu needed to ``expand its circle, monetize, and go public'' because the existing user base was insufficient for those goals [D06, C4]. Another described attracting male users as a way to diversify consumption categories and increase the value of advertising and ecommerce [D06, C37].

These claims represent users' interpretations of Xiaohongshu's strategy. The comments do not provide evidence about the platform's internal decision making. Their analytic significance lies in how users connected a highly visible content investment to their own experience of changing audience composition.

Evaluations of this perceived strategy varied considerably. Some commenters worried that growth would come at the expense of the existing community. One wrote, ``Do not replace all of us by the time you go public'' [D06, C18]. Another questioned the value of attracting male users if those users did not become consumers [D02, C145].

Others accepted or praised the commercial logic. One user described it simply as a business decision and wrote, ``Making money is nothing shameful'' [D02, C150]. Another argued,

\begin{quote}
\small
``Last year, the influx of foreign users brought in a large wave of male users. Now most of those foreign users are gone. Xiaohongshu clearly does not want the male users it had just attracted to disappear as well, so it quickly invested heavily in the World Cup rights to retain them and attract even more. Its strategic judgment is actually quite good.''

\hfill\textit{[D08, C197]}
\end{quote}

The disagreement was therefore less about whether audience composition could be actively shaped and more about how such growth should be evaluated. Critics and supporters alike frequently interpreted the World Cup through a model in which the platform could use major content investments to change who joined, who stayed, and which consumer groups became more important.

\subsection{Community Norms and Boundary Management}
\label{sec:findings-boundary-management}

Perceived audience change also made expectations about participation more explicit. This theme appeared in 76 comments across all eight source threads. Users discussed what kinds of behavior they considered acceptable, how disagreement should be expressed, and what should happen when others violated those expectations. These comments ranged from invitations to coexist under shared norms to blocking, reporting, confrontation, exclusion, and withdrawal.

\subsubsection{Users articulated conduct based expectations for participation}

Some comments framed acceptable membership primarily through conduct. One user wrote,

\begin{quote}
\small
``I think gender does not matter as long as people have reasonable values and know how to behave respectfully.''

\hfill\textit{[D03, C14]}
\end{quote}

This position did not require newcomers to share the same identity as existing users. It placed conditions on interaction. Respectful conduct became the criterion through which participation was accepted.

The same logic appeared when users tried to regulate discussion itself. In a thread about men downloading Xiaohongshu for the World Cup, the original discussion included an explicit request,

\begin{quote}
\small
``I did not make this post to start a fight. I just thought it was interesting that guys downloaded Xiaohongshu for the World Cup. Please keep the discussion civil.''

\hfill\textit{[D08, C1]}
\end{quote}

Other commenters expressed similar frustration with interactions that quickly became polarized. One wrote, ``I hope Xiaohongshu does not turn into Weibo, with people arguing and taking sides every day'' [D07, C8]. Another asked users to state their point clearly because vague provocative posts tended to generate large arguments in the replies [D07, C42].

These comments show that users were defining community boundaries through interactional expectations as well as identity. Conflict itself became an object of governance. Users discussed what counted as a legitimate disagreement, when a comment was unnecessarily provocative, and how much antagonism the community should tolerate.

Counterexamples also show that audience expansion was not uniformly experienced as a deterioration of norms. Some users welcomed the new participation. One commenter argued that the World Cup had brought in more ``normal people'' and allowed previously silent users to speak in discussions that had once felt one sided [D04, C15]. These comments complicate an account in which existing users uniformly defended a stable community against outsiders. Different users disagreed about the condition of the existing community and about whether newcomers improved or damaged it.

\subsubsection{Boundary management involved several different repertoires}

Among the 262 comments, 97 articulated an explicit boundary management or coping strategy. The most common primary strategies were algorithmic separation in 28 comments and content filtering in 24. Fifteen comments emphasized blocking or reporting. Eight emphasized gatekeeping, and another eight emphasized disengagement. Normative accommodation and norm education each appeared as the primary strategy in six comments, while direct confrontation appeared in two.

These categories capture different ways users located responsibility for maintaining a tolerable interaction environment. Some relied primarily on technical controls. Others tried to explain community expectations to newcomers. Some sanctioned individual users directly.

One commenter advised others,

\begin{quote}
\small
``There is one here. If you see them, block and report them. Do not argue too much.''

\hfill\textit{[D04, C27]}
\end{quote}

This response combines collective attention with platform provided enforcement tools. The user was not simply blocking an account for personal convenience. The comment instructed others in the thread on how to respond to a perceived disruptive participant.

Other comments combined explanation with sanction. One user described an ambiguous hostile comment that they believed could shape how other women interpreted the tone of a discussion. They wrote,

\begin{quote}
\small
``I explained what had happened clearly in the replies, and then blocked that person so they would not have space to keep speaking.''

\hfill\textit{[D04, C29]}
\end{quote}

Here, boundary work involved both interpretation and action. The commenter first attempted to establish how the problematic interaction should be understood, then used blocking to limit further participation.

Some comments were more openly confrontational. One user reported,

\begin{quote}
\small
``I already confronted them in a private message, blocked them, and reported them. Sisters, keep an eye out.''

\hfill\textit{[D04, C10]}
\end{quote}

The collective address is significant because the response was presented as something other users should notice and potentially participate in. Several comments from this source thread similarly coordinated blocking, reporting, or exclusion through explicitly collective language.

Withdrawal represented another response. Users sometimes concluded that maintaining their preferred environment required reducing participation or leaving. One commenter wrote,

\begin{quote}
\small
``The comment sections have become really unpleasant lately. If someone made a new text and image community, I would move there immediately.''

\hfill\textit{[D05, C16]}
\end{quote}

Another responded to the possibility of increasing male participation with, ``If there are more men, I will just leave. I have been a cyber nomad for a long time anyway'' [D03, C93].

These strategies should not be read as a sequence in which users first educate, then sanction, and finally leave. The comments do not provide temporal evidence for such an escalation. They show a repertoire of responses that coexisted in the discourse. Some users emphasized accommodation and civility. Others emphasized technical filtering, interpersonal sanction, exclusion, or exit. The variation illustrates disagreement over both what the community boundary should be and who should carry the work of maintaining it.

\subsection{Gendered Membership Was Contested}
\label{sec:findings-gender}

Gender was one of the most visible ways users described audience change, yet the meaning of gender for community membership remained disputed. This theme appeared in 104 comments across all eight source threads. Some users described Xiaohongshu's female skewed audience as part of the platform's existing identity. Others argued that gender should have little bearing on participation. A third set of comments connected the platform's apparently gendered character to differences in interests, use patterns, and personalized recommendation.

\subsubsection{Some users treated the existing gender composition as part of community identity}

Several comments assumed that the platform had developed a recognizable gendered character. Users discussed Xiaohongshu in comparison with platforms they perceived as more male dominated and questioned why its existing audience composition needed to change.

One commenter asked,

\begin{quote}
\small
``Every platform has a different gender mix. Why does it have to be changed?''

\hfill\textit{[D03, C110]}
\end{quote}

The question frames uneven gender composition as an ordinary feature of platform differentiation. From this perspective, a female skewed audience did not constitute a problem that required correction.

Some comments used stronger boundary language. Users described Xiaohongshu as a space whose existing culture might be disrupted by efforts to attract male audiences. Others interpreted the platform's growth strategy as a shift in who the platform valued. These claims sometimes overlapped with the commercialization narratives described in Section~\ref{sec:findings-growth}.

At the same time, other users challenged the premise that Xiaohongshu still had a fixed women centered identity. One commenter responded to a description of the platform as a ``young women's consumer decision community'' by asking who still used that label in 2026 [D02, C167]. Another wrote,

\begin{quote}
\small
``Xiaohongshu has always wanted to become China's Instagram. It is not some women only gym.''

\hfill\textit{[D03, C108]}
\end{quote}

These comments show that even users who recognized an uneven gender distribution disagreed about whether that distribution created any legitimate membership boundary.

\subsubsection{Users also defined belonging through behavior, interests, and everyday utility}

A recurring alternative was to detach membership from gender and define it through conduct. One commenter captured this position using Xiaohongshu's gendered language of solidarity.

\begin{quote}
\small
``As long as we do not turn everything into a gender issue, everyone can be sisters here. Men can be sisters too.''

\hfill\textit{[D08, C246]}
\end{quote}

The phrase ``sisters'' is ordinarily gendered, yet the commenter used it as an interactional category. A man who participated without turning discussion into gender conflict could be incorporated into the same collective label. The comment therefore preserved a recognizable community vocabulary while making its boundary conditional on behavior.

Explicitly male users also described long term participation that did not fit an invasion narrative. One wrote,

\begin{quote}
\small
``I am entirely in the study and exam part of Xiaohongshu. I have used it for two years and almost everything I see is study material. I have barely encountered gender topics, and I have never been excluded just because I am a man.''

\hfill\textit{[D03, C27]}
\end{quote}

This account connects gendered experience with both platform use and personalization. The user participated in a specific content environment and reported little exposure to the conflicts that dominated some of the sampled discussions.

Interests were similarly contested as evidence of gender. Some commenters assumed that football would primarily attract men. Others directly challenged that assumption. One wrote simply, ``Women users watch football too'' [D06, C9]. Another argued that the existence of women who watch football was fully compatible with men constituting a larger share of football audiences [D06, C13]. These exchanges show that users debated the relationship between gender categories and interests instead of taking that relationship as fixed.

Personalization offered another explanation for why Xiaohongshu could appear highly gendered to different users. One commenter wrote,

\begin{quote}
\small
``Xiaohongshu is basically a modern version of Tieba. There has never been a gender restriction. The information bubbles are just very strong, which makes it easy to find people with similar interests.''

\hfill\textit{[D03, C107]}
\end{quote}

This comment connects the gender debate back to the recommendation patterns described in Section~\ref{sec:findings-personalization}. A platform can appear socially homogeneous from within a highly personalized feed even while containing a more heterogeneous population.

Other commenters emphasized practical value. Users described relying on Xiaohongshu for travel information, study advice, product reviews, tutorials, and everyday problem solving. One explicitly male commenter said that Xiaohongshu ``solves many problems for us guys too'' [D07, C54]. Another long term male user jokingly asked whether he counted as an ``original male character'' on Xiaohongshu because he had joined years earlier as an alternative to Weibo [D01, C84].

These accounts complicate any simple division between established female users and newly arriving male users. Some male users described themselves as established participants. Some women rejected gender as a meaningful boundary. Some commenters welcomed new users under conduct based expectations, while others viewed the existing gender composition as worth protecting. The gender boundary in these discussions was therefore actively produced, challenged, and renegotiated through claims about history, behavior, interests, usefulness, and recommendation.

\subsection{Attribution and Boundary Management Strategies}
\label{sec:findings-attribution-strategy}

The qualitative findings showed substantial variation in both how users explained perceived community change and how they responded to it. We next examined whether these two dimensions were associated within individual comments. The primary analysis included 70 comments that contained both an explicit attribution covered by the three broader attribution families and an explicit boundary management strategy. The analysis therefore concerns comments in which both a problem explanation and a proposed or enacted response were visible.

\begin{figure}[t]
    \centering
    \includegraphics[width=\linewidth]{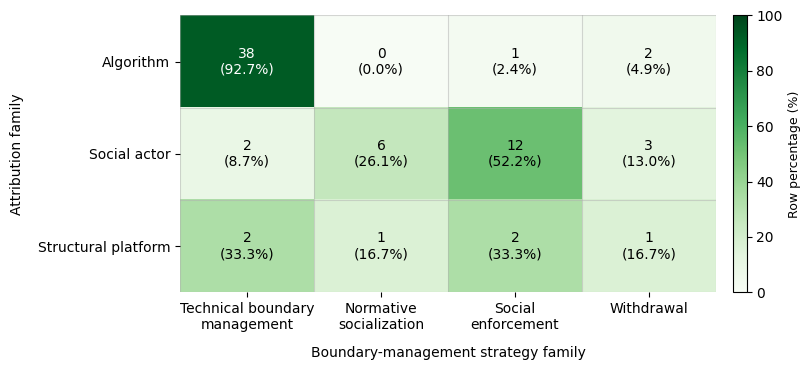}
    \caption{
    Distribution of broad boundary-management strategies across attribution families among comments containing both an explicit attribution and an explicit strategy ($N=70$). Percentages are calculated within attribution families. Algorithmic attributions were concentrated in technical boundary management, whereas social-actor attributions were more often associated with social enforcement and normative socialization. The permutation-based association test was significant ($p<.001$, Cram\'er's $V=.583$). The structural-platform category contains only six observations and should be interpreted cautiously.
    }
    \Description{
    A three-by-four heatmap showing boundary-management strategy distributions within three attribution families. Among 41 algorithmic attributions, 38, or 92.7 percent, were paired with technical boundary management. Among 23 social-actor attributions, 12, or 52.2 percent, were paired with social enforcement and six, or 26.1 percent, with normative socialization. The structural-platform row contains six comments distributed across all four strategy families.
    }
    \label{fig:attribution-strategy}
\end{figure}

The association between attribution and strategy was statistically significant. The permutation based test using 50,000 permutations produced $p < .001$, with Cram\'er's $V = .583$. Figure~\ref{fig:attribution-strategy} shows the distribution of strategy families within each attribution family.

The clearest pattern concerned algorithmic attribution. Among the 41 comments that attributed the perceived problem to the recommendation system, 38 comments, or 92.7\%, articulated technical boundary management. These responses involved relying on personalized separation or using content controls to reduce unwanted exposure. No comments in this group used normative socialization as their primary strategy. One used social enforcement, and two articulated withdrawal.

This concentration is consistent with comments in Section~\ref{sec:findings-personalization}. When users described recommendation as the source of unwanted exposure, their proposed response usually targeted recommendation as well. They clicked ``not interested,'' adjusted preferences, attempted to block unwanted topics, or relied on personalization to restore separation.

The distribution was different among social actor attributions. Of the 23 comments that attributed the perceived problem to incoming male users or community dynamics, 12 comments, or 52.2\%, articulated social enforcement. Six, or 26.1\%, articulated normative socialization. Three articulated withdrawal, and only two, or 8.7\%, articulated technical boundary management. Comments in this group therefore more often located both the problem and the response within interaction between users.

The structural platform category contained six comments with an explicit strategy. Two articulated technical boundary management, two articulated social enforcement, one articulated normative socialization, and one articulated withdrawal. The small number of observations makes this internal distribution difficult to interpret, so we do not draw a substantive conclusion from differences within this category.

We also examined whether this association depended heavily on one of the eight source discussions. Repeating the analysis while excluding each source in turn produced Cram\'er's $V$ values ranging from .561 to .626. The association therefore remained similar in magnitude when any single source was removed.

Finally, we examined whether attribution was also associated with evaluative stance. Among the 173 comments containing an explicit attribution, the permutation based test did not detect a statistically significant association between attribution and stance, with $p = .182$ and Cram\'er's $V = .200$. The stronger pattern in our data concerned what users proposed doing in response to different perceived causes. Users who located the problem in different parts of the platform and community did not clearly differ in whether they evaluated the change positively or negatively, while their articulated responses showed substantially greater variation.


\section{Discussion}
\label{sec:discussion}

Our findings connect perceived audience change to four issues: growth as a condition of community governance, recommendation as infrastructure for informal social boundaries, causal sensemaking as part of boundary management, and gendered membership as negotiated community identity. Across these themes, users were negotiating both what had changed and where responses should be directed.

\subsection{Platform Growth as Community Governance}
\label{sec:discussion-growth}

Users frequently interpreted Xiaohongshu's World Cup investment as an audience growth strategy, particularly as an effort to attract or retain male users. These comments do not provide evidence about Xiaohongshu's internal goals or decision making. Their significance lies in how users understood the intervention. A content investment was interpreted as changing the population of the platform and, through that change, the social conditions under which participation occurred.

Prior research shows that changes in participation can make community norms newly consequential. Kiene et al.'s study of NoSleep found that a sudden newcomer influx created a governance challenge managed through coordinated administration, community enforcement, and technical mechanisms \cite{kiene2016eternal}. Their later study of World of Warcraft guild mergers further shows that the consequences of membership change depend on the practices and social relationships that newcomers bring \cite{kiene2018mergers}. Our case extends this work to a less formally bounded setting. Xiaohongshu does not organize the broader platform community through explicit membership in the way that a subreddit or guild does. As Form-From helps clarify, platforms may organize participation through spaces, networks, or a broader commons, producing different relationships between content distribution and community structure \cite{zhang2024formfrom}. In our data, audience change became consequential even though there was no formal platform-wide community boundary for newcomers to cross.

This distinction helps refine what it means to describe platform growth as a governance issue. Prior work distinguishes community self-moderation from centralized corporate moderation, while multi-level governance research locates authority across platform operators, intermediate governance structures, and end users \cite{seering2020selfmoderation,jhaver2023decentralizing}. Recent work on community appeals further shows how formal community-level procedures can occupy this intermediate layer \cite{lee2025appeals}. Our findings draw attention to a decision that can occur before familiar forms of moderation or rule enforcement become necessary. When a platform-level intervention changes who participates or becomes visible, it can also change the population across which existing norms must be interpreted, communicated, and enforced. In the sampled discussions, questions about civility, appropriate participation, blocking, gatekeeping, and withdrawal were discussed alongside perceptions of audience growth.

Wan et al. document a closely related dynamic on Xiaohongshu. Their participants connected hashtag-based audience control to a shifting user base, perceived attempts to attract male users, gender-based harassment, and dissatisfaction with existing safety tools \cite{wan2025hashtag}. Our results confirm that platform strategy and audience composition are already linked in users' understandings of Xiaohongshu. They extend this finding from a specific audience-control practice to public discussion of a platform-wide audience shift. The contribution is therefore not that growth decisions are equivalent to formal moderation. It is that audience acquisition can create governance consequences by altering the social environment in which norms, membership claims, and user-level governance practices operate.

For HCI, this suggests that research on platform governance should attend to interventions that change participation before conflict is expressed as a conventional moderation problem. Major content partnerships, promotional campaigns, and other audience-acquisition efforts can become socially meaningful when users perceive them as changing who constitutes the platform community. The design problem is partly one of transition management: supporting existing and incoming users as expectations about participation become more visible and potentially contested.

\subsection{Recommendation Systems as Social Boundary Infrastructure}
\label{sec:discussion-boundary-infrastructure}

Our findings also show that users understood personalization in social terms. Some commenters described highly personalized feeds as allowing different interests and social circles to coexist with limited contact. Others described repeated encounters with unwanted or contentious content after attempting to filter it. The same recommendation system could therefore be experienced as maintaining separation or as failing to preserve it.

Existing research already shows that recommender systems can shape community formation and social connection. Wang et al.'s interviews with Xiaohongshu users who had experienced mental illness describe an ``algorithmically woven community'' in which recommendation connected similar users across a loosely knit and decentralized network \cite{wang2023woven}. Zhao similarly found that long-term gay male Zhihu users could understand personalization as protective, enabling reduced contact with hostile outsiders and supporting relatively exclusive networks \cite{zhao2023filter}. These studies demonstrate that algorithmic curation has social consequences beyond selecting relevant content.

Our findings extend this work by focusing on users' experience of separation between heterogeneous audiences. The relevant question in our case was not only whether recommendation helped similar users find one another. Commenters also expected personalization to keep some topics, interests, and contentious interactions outside their ordinary feed. This expectation became visible when it failed. Marking a post as ``not interested'' or attempting to retrain a feed could therefore function as boundary repair. The goal expressed in these comments was often to restore a familiar content environment in which unwanted encounters became less frequent.

This pattern also qualifies accounts in which permeability is primarily associated with discovery and connection. Milton et al.'s study of TikTok mental-health communities shows that algorithmically mediated communities can be highly permeable, enabling users to encounter new communities while also creating situations in which harmful or unwanted content becomes difficult to escape \cite{milton2023iseeme}. Van der Nagel similarly theorizes ``forced connections'' to describe situations in which algorithmic systems connect otherwise disparate people, content, or data, prompting users to develop explanations and tactics for intervening in those connections \cite{vanderNagel2018networks}. Our case shows a related process at the level of perceived platform communities. Users sometimes interpreted unexpected exposure as evidence that previously separate social environments had become more porous.

We use \textit{social boundary infrastructure} to describe this role of recommendation. The term does not imply that recommender systems create fixed social groups or that personalization objectively separates demographic populations. It identifies a user-experienced function of personalized distribution. Repeated recommendation decisions can shape which audiences users regularly encounter, which remain largely outside everyday experience, and how dependable that separation appears. When users rely on this infrastructure, failures of personalization may be experienced as changes in social proximity as well as failures of content relevance.

This framing also helps distinguish our findings from Wan et al.'s audience-control case. Wan et al. show users manipulating recommendation cues to reduce the probability that unwanted male audiences would receive their posts \cite{wan2025hashtag}. Our commenters more often discussed the receiving side of recommendation, including feed separation, ``not interested'' feedback, and repeated unwanted exposure. Together, these findings point to complementary creator- and consumer-side boundary needs on recommendation-driven platforms.

\subsection{Causal Sensemaking and User Boundary Management}
\label{sec:discussion-sensemaking}

RQ1 showed that users distributed explanations of perceived community change across several parts of the sociotechnical environment. RQ2 showed that their proposed responses were similarly distributed across technical controls, normative interaction, interpersonal enforcement, and withdrawal. The quantitative analysis connects these findings. Among comments containing both an explicit attribution and an explicit strategy, attribution and strategy were strongly associated, with Cramer's $V=.583$. Algorithmic attributions were overwhelmingly paired with technical boundary management, while social-actor attributions were much more often paired with social enforcement or normative socialization. The association remained similar in magnitude when each source thread was excluded in turn.

We interpret this pattern as an alignment between the perceived location of a problem and the perceived site of intervention. This interpretation remains non-causal. A comment may articulate an explanation because a particular response already seems appropriate, a preferred response may follow from an explanation, or both may be produced together through a broader understanding of the situation. Our data cannot distinguish these possibilities.

Prior folk-theory research provides a useful basis for understanding this relationship. During the \#RIPTwitter controversy, DeVito, Gergle, and Birnholtz found that more elaborated theories of a proposed algorithmic change tended to accompany more specific reactions \cite{devito2017algorithms}. Mayworm et al. similarly show that marginalized users' folk theories of content moderation, including broader perceptions of platform spirit, can inform changes in language, reduced participation, and platform exit \cite{mayworm2024moderation}. These studies establish that user explanations can be practically consequential. Our findings extend this relationship from explanations of algorithmic or moderation outcomes to explanations of perceived community-level change.

Zhang and Yang provide an especially important comparison because their analysis of public Weibo and Twitter discourse shows that lay explanations of information homogeneity are distributed across groups, individuals, political actors, platforms, and algorithms \cite{zhang2025beyond}. Their results caution against assuming that users understand sociotechnical outcomes primarily through algorithms. Our corpus shows a similar multiplicity. Users attributed perceived community change to recommendation, platform and commercial decisions, incoming users, and community dynamics. We extend this public-discourse perspective by examining how these different causal locations correspond to different forms of articulated boundary intervention.

The secondary stance analysis sharpens this interpretation. We did not detect a statistically significant association between attribution and evaluative stance among comments with an explicit attribution. Users who located change in different parts of the sociotechnical system were therefore not clearly separated by whether they viewed the change positively or negatively. The more pronounced relationship concerned what they proposed doing about it. In this corpus, causal sensemaking appears especially relevant to the allocation of responsibility and intervention.

This also helps explain why boundary management took several forms. Technical problem framings made filtering and personalization relevant, while social explanations made norm communication, blocking, reporting, and gatekeeping more salient. Formal moderators contribute to community development through rules and moderation practices \cite{seering2019moderator}; our corpus instead shows users articulating boundary work without a recognized moderator role. Moderation explanations can also affect bystanders \cite{jhaver2024bystanders}, providing a formal analogue for the public norm communication in our data. Collective encouragement to block or report likewise resembles prior user-led governance \cite{geiger2016blocklists,burrell2019control}. Withdrawal need not mean binary exit: Facebook non-use research documents degrees of limiting, leaving, and reconnecting \cite{baumer2013limiting}. Our repertoire similarly treats disengagement as one possible boundary-management response.

For design, the implication is that a single class of control is unlikely to address every way users understand community disruption. Platforms can support several intervention sites, including intelligible content controls, effective blocking and reporting, and mechanisms for communicating interaction expectations. These controls should remain available without requiring the system to infer which causal explanation a user holds. The finding concerns the diversity of perceived problems and remedies, not a basis for automatically classifying users into intervention types.

\subsection{Gendered Membership as Negotiated Community Identity}
\label{sec:discussion-gender}

Gender was a prominent vocabulary through which users described Xiaohongshu's history and perceived audience change, but it did not provide a stable membership rule. Some comments treated the platform's female-skewed history as meaningful to community identity and viewed increased male participation as potentially disruptive. Others explicitly rejected gender as a criterion for belonging and emphasized conduct, shared interests, or everyday platform use. In some comments, even gendered language such as ``sisters'' could include men who were perceived as participating appropriately.

Prior work shows that governance and participation can contribute to platform identity and identity-based boundaries. Das et al. show how sociotechnical governance can privilege some identities while marginalizing others on Bengali Quora \cite{das2021jol}, while Beadle et al. show implicit norms being created, contested, and reinforced in non-binary Reddit communities \cite{beadle2025implicit}. Ammari et al.'s women-only Facebook groups offer a useful contrast because membership boundaries there were explicit and actively maintained through moderation \cite{ammari2022moderation}. Our case concerns a broader setting where users disputed whether gender should define the boundary at all.

Qin et al. similarly show that gender can become a contested interpretive frame in Chinese social-media discourse \cite{qin2024genderblindness}. A related disagreement appeared in our corpus at the level of community membership: some users treated gender composition as central, while others relocated the problem to conduct, interests, recommendation, or antagonistic interaction. Haapoja et al.'s work on ideal discussants and audiences provides a useful parallel by showing how participation can be evaluated through conduct and role expectations \cite{haapoja2025ideal}. Our findings extend this issue to contested platform membership, where conduct and interests sometimes displaced gender as the basis for belonging.

This finding also refines the closest Xiaohongshu precedent. Wan et al. document women users deliberately attempting to reduce male audience exposure in response to harassment, safety concerns, and perceived changes in the platform's user base \cite{wan2025hashtag}. Our data confirm that gendered audience change and platform identity are salient concerns on Xiaohongshu. They also show a broader range of positions within event-focused public discourse. Some users defended a gendered understanding of the community, while others accepted male participation under conduct-based expectations or disputed the premise that gender should determine membership.

We therefore treat gendered platform identity as negotiated rather than as a demographic fact that can be inferred directly from audience composition. In this case, gendered membership was articulated through historical participation, perceived platform strategy, recurring content practices, recommendation-mediated exposure, and interactional norms. For HCI research, this suggests that studies of identity-based platform communities should examine how users themselves make membership criteria consequential. A female-skewed audience, a history of women-centered content, and users' gendered descriptions of a platform may all matter, but they do not by themselves establish a universally shared community boundary.

\subsection{Design Implications}
\label{sec:discussion-design}

Our findings suggest several design considerations for recommendation-driven platforms experiencing visible audience change. These are design opportunities, not interventions validated by this study.

First, recommendation controls might be understood as mechanisms for \textit{boundary repair}, not only signals of content relevance. Commenters relied on personalization to keep interests and social circles relatively separate, while repeated unwanted exposure made that separation feel unreliable. Existing controls such as ``not interested'' therefore carried a social as well as content-selection function. One direction is to make the scope and consequences of such feedback more legible, for example by distinguishing preferences concerning a topic, event, or recurring source and indicating whether those preferences affect subsequent recommendation. This extends creator-side audience-control concerns identified by Wan et al. \cite{wan2025hashtag} and aligns with work emphasizing individualized control rather than uniform definitions of unwanted content \cite{rashed2026transformation}. Stronger boundary controls, however, should neither assume separation is always desirable nor infer demographic boundaries on users' behalf.

Second, platforms may need to support multiple intervention sites. Attribution and strategy were strongly associated: users who located disruption in recommendation overwhelmingly articulated technical responses, whereas social-actor attributions more often accompanied normative or enforcement-based responses. Rather than treating filtering, moderation, and norm communication as interchangeable, platforms might clarify what each intervention acts upon. Seeking less exposure to an event differs from reporting harassment or communicating an interaction norm. The goal is to make these pathways intelligible without inferring or classifying users by an assumed causal model.

Third, audience transitions create tension between supporting established expectations and institutionalizing contested informal boundaries. Users disagreed over whether gender, conduct, interests, platform history, or everyday use should define legitimate membership. This contrasts with settings where identity-based membership is explicitly established and moderated \cite{ammari2022moderation}. Our findings therefore favor conduct-based, revisable norm communication over mechanisms that automatically encode demographic membership criteria. Contextual guidance or newcomer-facing expectations may make interaction norms more visible, but should remain open to contestation rather than convert one group's community identity into a platform-wide rule.

Finally, large audience-acquisition initiatives may warrant attention as social transitions as well as growth interventions. Our data do not establish that the World Cup investment caused platform-wide demographic or normative change, but users interpreted audience growth alongside renewed negotiation over participation and belonging. Platforms might therefore consider how major content partnerships or promotional events alter encounters between existing and incoming users. The design challenge is to support continuity, newcomer participation, and user agency without assuming that either existing norms or increased cross-group exposure should automatically take precedence.

\subsection{Limitations and Future Work}
\label{sec:limitations}

This study examines eight event-focused discussion threads and 262 analytic comments. The corpus was constructed around discourse concerning a perceived World Cup related audience influx and should not be treated as representative of Xiaohongshu users or platform-wide discussion. Our purposive selection of relatively high-engagement posts may overrepresent contexts in which audience change or conflict was especially salient. Platform ranking and recommendation may also have influenced which source posts and comments became visible during collection. Future work could compare event-focused discourse with broader samples collected across topics, events, and time periods.

We could not reliably determine the gender of most commenters and did not infer gender from usernames, avatars, interests, or writing style. Gender-specific interpretation was limited to explicit self-identification or explicit collective positioning when analytically relevant. Our findings therefore concern gendered discourse and contested membership within the sampled discussions. They do not estimate attitudes among women or men on Xiaohongshu as demographic populations. Interviews, surveys, or opt-in demographic data could examine how people with different identities and histories of platform use experience audience change.

The study also analyzes publicly articulated explanations and strategies. A comment may report an action, recommend one to others, or state a position without revealing the commenter's full beliefs or behavior outside the observed interaction. A strategy code of \textit{none} means that no strategy was explicit in the comment, not that the user had no response. Likewise, users' attributions capture their explanations of platform and community change and do not reveal Xiaohongshu's internal decisions or recommendation mechanisms.

Finally, the association between attribution and strategy is observational and should not be interpreted causally. Comments were nested within eight source threads, creating source-specific dependence in comment-level patterns. The leave-one-source-out analysis suggests that the primary association was not driven by a single discussion, but it does not remove the limitation of the small number of source threads. The cross-sectional corpus also does not establish temporal ordering among different forms of boundary management. In particular, our data do not support an escalation sequence from accommodation to stronger enforcement or withdrawal. Longitudinal research, broader source sampling, interviews, and platform-level data could examine how causal explanations and boundary-management practices develop over time.

\section{Conclusion}

This study examined public Xiaohongshu discussions around a perceived World Cup-related audience influx to understand how users attributed community change and negotiated boundaries. Across 262 comments, users described personalization as both social separation and a boundary that could fail, interpreted the World Cup investment through audience growth and commercialization, and disputed whether gender should define membership. Attribution was strongly associated with articulated strategy: algorithmic attributions concentrated in technical boundary management, whereas social-actor attributions more often aligned with normative or enforcement responses. These findings show how audience change on recommendation-driven platforms is interpreted through technical and social frames, linking causal sensemaking to where users imagine intervention should occur.

\begin{acks}

\end{acks}

\bibliographystyle{ACM-Reference-Format}
\bibliography{references}

\appendix


\end{document}